%% file: main.tex
\documentclass[10pt,conference]{IEEEtran}

\usepackage{xspace}
\usepackage{framed}
\usepackage{graphicx} 
\usepackage{makecell}
\usepackage{subfigure}
\usepackage{enumitem}
\usepackage[hyphens]{url}
\usepackage{breakurl}
\usepackage[numbers,sort&compress]{natbib}
\usepackage{color}
\usepackage{authblk}

\newcommand{\sysname}{Elevate\xspace}

\title{Model-Enhanced LLM-Driven VUI Testing of VPA Apps}
\author[1]{Suwan Li}
\author[1]{Lei Bu}
\author[2]{Guangdong Bai}
\author[2]{Fuman Xie}
\author[3]{Kai Chen}
\author[3]{Chang Yue}
\affil[1]{Nanjing University}
\affil[2]{University of Queensland}
\affil[3]{Institute of Information Engineering, Chinese Academy of Sciences}

\begin{document}
\maketitle
\input{ICSE2025/abstract}

\input{ICSE2025/introduction}

\input{ICSE2025/background}

\input{ICSE2025/overview}

\input{ICSE2025/evaluation}

\input{ICSE2025/discussion}

\input{ICSE2025/related-work}

\input{ICSE2025/conclusion}


\bibliographystyle{IEEEtran}
\bibliography{IEEEabrv,reference}
\end{document}

%% file: ICSE2025/abstract.tex
\begin{abstract}
The flourishing ecosystem centered around voice personal assistants~(VPA), such as Amazon Alexa, has led to the booming of VPA apps.
The largest app market Amazon skills store, for example, hosts over 200,000 apps. 
Despite their popularity, the open nature of app release and the easy accessibility of apps also raise significant concerns regarding security, privacy and quality. 
Consequently, various testing approaches have been proposed to systematically examine VPA app behaviors. 
To tackle the inherent lack of a visible user interface in the VPA app, two strategies are employed during testing, i.e., chatbot-style testing and model-based testing.
The former often lacks effective guidance for expanding its search space, while the latter falls short in interpreting the semantics of conversations to construct precise and comprehensive behavior models for apps.

In this work, we introduce \sysname, a model-enhanced large language model~(LLM)-driven VUI testing framework.
\sysname leverages LLMs' strong capability in natural language processing to compensate for   semantic information loss during model-based VUI testing.
It operates by prompting LLMs to extract states from VPA apps' outputs and generate context-related inputs.
During the automatic interactions with the app, it incrementally constructs the behavior model, which facilitates the LLM in generating inputs that are highly likely to discover new states. 
\sysname bridges the LLM and the behavior model with innovative techniques such as encoding behavior model into prompts and selecting LLM-generated inputs based on the context relevance. 
\sysname is benchmarked on 4,000 real-world Alexa skills, against the state-of-the-art tester Vitas. 
It achieves 15\% higher state space coverage compared to Vitas on all types of apps,  
and exhibits significant advancement in efficiency.
\end{abstract}

%% file: ICSE2025/introduction.tex
\section{Introduction}
With the prevalence of smart speakers, voice personal assistants (VPA) have permeated various aspects of people's lives.
Prominent examples include Amazon Alexa, Google Assistant, and Apple Siri, which have been widely used for assisting smart speaker users. 
Centered around them, numerous applications (or VPA apps for short) have been developed to provide various functionalities, such as accessing news, entertainment, and controlling devices. 
VPA apps are characterized by the \emph{voice user interface} (VUI), which enables user interaction solely through verbal conversations. 

The major VPA service providers have established VPA app stores for efficient app distribution. 
Through them, third-party developers can unload their apps, and users can invoke apps without installation, simply by calling their invocation names.
Such openness and ease of access have led to the widespread popularity of VPA apps.
For example, the skills store, the largest VPA app store, boasts over 200,000 apps~\cite{skill_number}.
However, there have been concerns raised regarding their security, privacy and quality.
A considerable number of VPA apps are found malicious as a result of untrustworthy skill certification process~\cite{dangerous_skill_in_store,skill_security}. 
Prior works have discovered that malicious VPA apps can eavesdrop~\cite{alexa_listening,skill_record_conversations} or ask users' privacy information without permissions~\cite{skillExplorer,skipper}.
The behavior of several VPA apps contradicts their privacy policies~\cite{skillExplorer,skipper,skilldetective,verhealth}.
Additionally, a large number of apps exhibit poor quality, such as terminating unexpectedly~\cite{vitas} or failing to understand common user inputs~\cite{LipFuzzer}.

\begin{figure}[t]
    \centering
    \subfigure[Semantic relevant inputs should have higher priority.]{
        \begin{minipage}{0.8\linewidth}
            \centering
            \includegraphics[width = 0.9\textwidth]{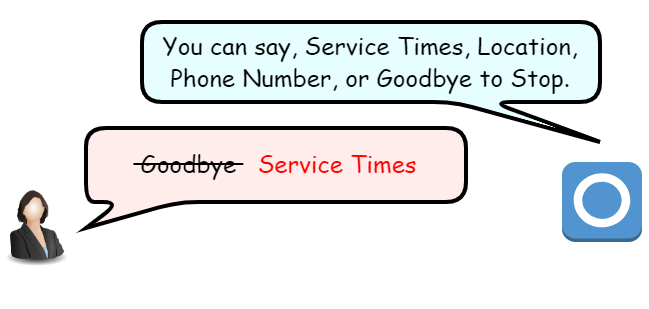}
            \label{fig:example_1}
        \end{minipage}
    }
    \quad
    \subfigure[Semantic similar states should be merged.]{
        \begin{minipage}{0.8\linewidth}
            \centering
            \includegraphics[width = \textwidth]{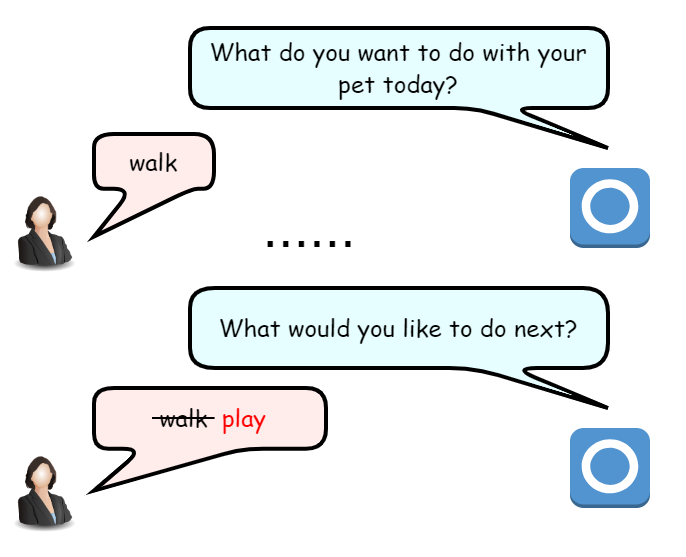}
            \label{fig:example_2}
        \end{minipage}
    }
    \centering
    \caption{Lack of semantic information impacts the testing efficiency.}
    \label{fig:motivating_example}
\end{figure}

To detect such problems, a thorough exploration of VPA apps' behavior is necessary.
Existing methods mainly employed strategies of depth-first search based chatbot-style testing~\cite{skillExplorer,VUI_UPSET_1,VUI_UPSET_2,skilldetective,verhealth} or model-based testing (MBT)~\cite{vitas}.
Since VPA apps cannot roll back to the previous interface, the exploration efficiency can be affected especially when the depth-first search strategy is taken.
Such testers have to start from the beginning after searching one path, resulting in repeated tests.
They can work effectively on simple apps, but may suffer from low efficiency when facing complex apps.
In addition, previous MBT approach falls short in understanding and utilizing semantic information when exploring apps' behavior and constructing the model.

Figure \ref{fig:motivating_example} shows two communication logs that illustrate the impact of semantic information on efficiently testing VPA apps.
In figure~\ref{fig:example_1}, between the candidate inputs ``Goodbye'' and ``Service Times'', ``Service Times'' is more likely to lead to unseen app behavior.
Therefore, ``Service Times'' should have higher initial priority than ``Goodbye''.
Without considering the semantic relevance of inputs, it is likely that ``Goodbye'' is selected and the app stops.
In figure~\ref{fig:example_2}, the two apps' outputs represent similar functional semantics but are expressed differently.
The user inputs ``walk'' at the first time, so other inputs like ``play'' should have higher priority at the second time.
However, if different outputs are considered as different functionalities, purposes or context, the same input ``walk'' will be selected at the second time for thorough testing.
The ignorance of outputs' semantic similarity at the level of functionality, purpose and context causes repeated tests.

Therefore, the semantic information is crucial in efficient testing of VPA apps.
As the large language models (LLM) are known for their strong natural language understanding and processing abilities~\cite{LLM_NLP_1,LLM_NLP_2,LLM_NLP_3,cot}, and previous studies have found that they can be used for downstream tasks with in-context learning~\cite{in-context-learning-1}, we adopt the LLM to drive the testing process to compensate for semantic information loss during the model-based VUI testing.
However, employing the LLM for the VUI testing presents the following three challenges:

\begin{figure}[h]
    \centering
    \includegraphics[width=0.4\textwidth]{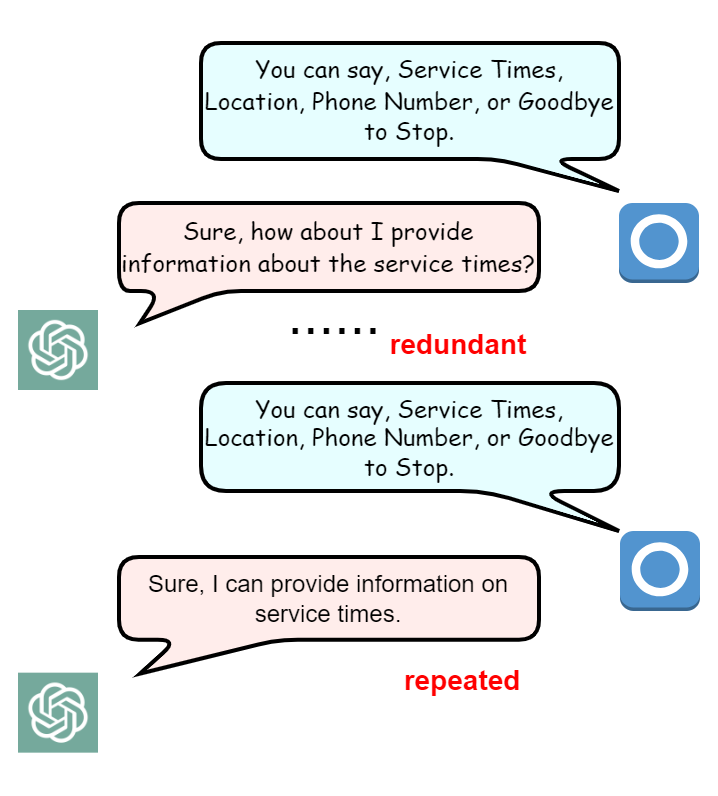}
    \caption{LLMs can generate redundant and repeated results if prompts are not carefully designed.}
    \label{fig:challenge}
\end{figure}

\noindent\textbf{Challenge 1}: 
LLMs can be used to supplement the semantic loss during the model-based testing of VPA apps, but it is difficult for LLMs to maintain the state information of VPA apps accurately.
On the one hand, when the testing goes deeper and the context becomes larger than the LLM's limitation, the information required for LLMs to generate an accurate model is incomplete.
On the other hand, LLMs can hardly generate a precious model especially when the VPA apps' behavior is complex.
However, a wrong model can greatly affect the following exploration.


\noindent\textbf{Challenge 2}: The results generated by LLMs can be redundant and repeated under VPA apps' context.
For example, if the LLM is asked to generate context-related inputs for a given VPA apps' outputs (see figure~\ref{fig:challenge}), it tends to generate long results, but most VPA apps have difficulty processing these inputs.
If state information and exploration strategy is not provided, the LLM can generate repeated inputs for the same state, affecting the testing efficiency.
For these reasons, prompts should be carefully designed to help the LLM generate formalized and efficient results.

\noindent\textbf{Challenge 3}: LLM's results are not entirely reliable due to its unexplainability and uncertainty.
For example, even if LLMs are prompted to return simple and concise results, they may still generate results that VPA apps cannot understand.
Therefore, we need to filter out the unreliable results based on the feedback from VPA apps and our domain knowledge.

To address the above three challenges, we propose the following solutions.

To tackle \textbf{Challenge 1}, we split the complex LLM-driven model-based testing tasks into three phases: states extraction, input events generation, and state space exploration to increase the accuracy of model construction.
In each phase, the LLM only extracts the state and generate input events for the real-time VPA apps' output, so the length of prompt will not exceed the context limitation.
Besides, the LLM is only used to make up for the semantic loss during the model construction and exploration, such as merging outputs with similar semantics to one state, generating context-related inputs and selecting an input for efficient exploration, while the model information is stored and maintained locally.

For addressing \textbf{Challenge 2}, we embed the information provided by the behavior model into the prompts to help the LLM generate efficient results and avoid repeated tests.
Since the complete behavior model is complex and occupies many tokens, adding it to the prompt not only interferes with the extraction of core information but also brings unnecessary expenses.
Therefore, we only extract phase-specific information to the prompt.
For example, only the state list is provided in the states extraction phase.
Meanwhile, by designing appropriate few shots, we enable the LLM to formalize outputs.
For the state space exploration, we implement the step-by-step chain-of-thought strategy to guide the LLM in parsing the behavior model and making decisions.

To handle \textbf{Challenge 3}, we establish specific rules considering both the behavior model information and VPA apps' feedback to check whether the LLM's outputs at each phase meet our requirements.
If they do not pass the checks, we provide feedback prompts for LLMs to regenerate the results.

Based on these ideas, we develop the \sysname(model-Enhanced Llm drivEn Vpa App's vui TEsting) framework.
As a model-based testing method, the \sysname framework is divided into three phases: states extraction, input events generation, and state space exploration.
These phases are enhanced by the LLM to achieve accurate state extraction and efficient state space exploration.
In the states extraction phase, the LLM is prompted to merge the VPA app's outputs with existing states in the behavior model or create a new state.
In the input events generation phase, the LLM generates context-related input events based on VPA app's outputs.
The states and input events generated by the LLM are used to update the behavior model.
Throughout the state space exploration process, the current-state related information from the behavior model is extracted and used to guide the LLM to select an input event for efficient exploration.

\noindent Our \textbf{contributions} are summarized as follows:
\begin{itemize}[leftmargin=*]
    \item We propose to use the LLM to enhance the model-based testing of VPA apps.
    This approach combines the model guidance of MBT with the NLP capabilities of the LLM.
    The LLM's results are used for constructing accurate behavior models and efficiently exploring the state space.
    \item We present a specific feedback mechanism to filter the LLM's unreliable results and guide LLMs for corrections.
    Based on the behavior model information and VPA apps' outputs, we filter out mismatched states, invalid input events and inefficient exploration strategies.
    \item We implement \sysname, and validate its coverage, efficiency, and generality.
    It surpassed the state-of-the-art approach Vitas in state space coverage and efficiency.
    Ultimately, \sysname tests 4,000 Alexa skills and covers 15\% of more state space than Vitas.
\end{itemize}

%% file: ICSE2025/background.tex
\section{Background}\label{sec:behavior_model}

\subsection{VPA Apps and Behavior Model}
VPA apps are apps based on smart speakers.
Users interact with VPA apps through voice, so the interface of VPA apps is called the voice user interface~(VUI).
VUIs are typically free of visible graphical interfaces.
Therefore, the exchange of all information are purely through voice.
While the VUI brings convenience, its invisible feature introduces a range of quality and security concerns, such as unexpected exits~\cite{vitas}, privacy violations~\cite{skill_security,alexa_listening}, and expected apps started~\cite{skill_squatting, skill_record_conversations}.
For this reason, thoroughly exploring VPA apps' behavior while testing the VUI's quality and security issues is of paramount importance.

However, VPA apps are not open source for normal testers.
A VPA app is composed of the front-end interaction model and the back-end processing code.
The development platform provides storage for the front-end interaction model, while the back-end code of VPA apps is stored on the developer's server.
As a result, dynamic testing is a commonly used method for testing the VUI of VPA apps.
Since the front-end interaction model of VPA apps is designed based on implicit models~\cite{action_model}, we propose to use the model-based testing approach to explore the behavior of VPA apps.

VPA apps' outputs express their functionalities and purposes.
By understanding and analyzing the outputs, states can be extracted.
Apps' transfer from one state to another is only triggered by users' inputs.
As a result, VPA apps' behavior can be described by the finite-state machine~(FSM), which has been proved to be applicable for constructing VPA apps' behavior models~\cite{vitas}.
A finite-state machine consists of five parts, described as $FSM = (Q, \Sigma,\delta,s_0,F)$. Among them:
\begin{itemize}[leftmargin=*]
    \item $Q$ represents the set of states. Apps' outputs are mapped to states.
    \item $\Sigma$ represents the set of input events. Users' inputs are mapped to input events.
    \item $F$ is the set of final states, and satisfies $F \subseteq Q$. VPA apps' final outputs are mapped to final states.
    \item $s_0$ is the initial state and satisfies $s_0 \in Q$. The initial state is always set as ``\textless START\textgreater''.
    \item $\delta: Q\times \Sigma \to Q $ represents a transition function. 
    The input event $e$ that triggers the transition from the state $s_0$ to the states $s_1$ is represented as $\delta(s_0, e) = s_1$.
\end{itemize}


\subsection{Large Language Model}
Large Language Model~(LLM) is built on the transformer architecture.
LLMs have been proved with strong natural language processing capabilities~\cite{LLM_NLP_1,LLM_NLP_2,LLM_NLP_3,cot}.
Compared to general language models~(LM), LLMs have a vast number of parameters and undergo extensive text training.
Due to these characteristics, LLMs can be directly applied to downstream tasks.
In addition, methods like fine-tuning~\cite{fine-tuning} and in-context learning~\cite{in-context-learning-1,in-context-learning-2} can improve LLM's capabilities for specific downstream tasks.
In the in-context learning technique, users only need to provide few samples as a reference for the downstream task, which implies that LLMs can handle downstream tasks through learning from a small dataset.

LLMs can be categorized into three types based on the transformer architecture: encoder-only, encoder-decoder, and decoder-only.
Encoder-only and encoder-decoder are suitable for infilling tasks, while decoder-only models are better at text generation tasks.
Considering that our tasks involve the model generation and exploration, we prefer to adopt decoder-only models.
Popular decoder-only models include OpenAI's GPT series~\cite{chatGPT,GPT4}, Meta's Llama series~\cite{Llama2}, etc.
Additionally, there are models specifically designed for code generation tasks such as Codex~\cite{codex} and Codegen~\cite{codegen}.

%% file: ICSE2025/overview.tex
\section{LLM Driven Model Construction and Exploration}

\subsection{Overview}
As a model-based testing framework, \sysname works by constructing the model according to VPA apps' behavior and guiding the exploration based on this model.
The behavior model is built by mapping VPA apps' outputs to states and users' inputs to input events (see Section~\ref{sec:behavior_model}).
As states reflect VPA apps' functionalities, purposes and behavior, different outputs with similar semantics (e.g., functionalities, purposes and behavior) should be mapped to one state.
We call these outputs as semantically similar outputs under the context of VPA apps' behavior.
Besides, users' inputs should be context related to the apps' outputs so that meaningful states can be discovered.
Overall, the states extraction and input events generation require natural language processing, which is the strength of the LLM.

In addition, the LLM has proved its ability in understanding graphs~\cite{GPT4} and reasoning with prompt engineering techniques such as in-context learning and chain-of-thought~\cite{in-context-learning-1,in-context-learning-2,cot}.
Our state space exploration task is basically an input event selection task considering factors like historical transitions, invocation frequency and relevance to the current state based on understanding the behavior model (i.e., a graph).
Given current state related information from the behavior model, the LLM can be used to select input events for further exploration of VPA apps' behavior.

In traditional model-based testing, the model is firstly built and then used to guide the exploration of the state space.
However, when testing VPA apps, the initial model is difficult to acquire before interacting with VPA apps as the VPA apps are closed-source and most documents only provide a few lines to describe their functionalities.
To solve that problem, we construct VPA apps' behavior model on-the-fly, which means the model is built during the interaction.
The behavior model is finally embedded into the prompt to guide the LLM in extracting states and selecting efficient input events for exploration. 
To save tokens, only phase-specific behavior model information is provided.

\begin{figure*}
    \centering
    \includegraphics[width=0.95\textwidth]{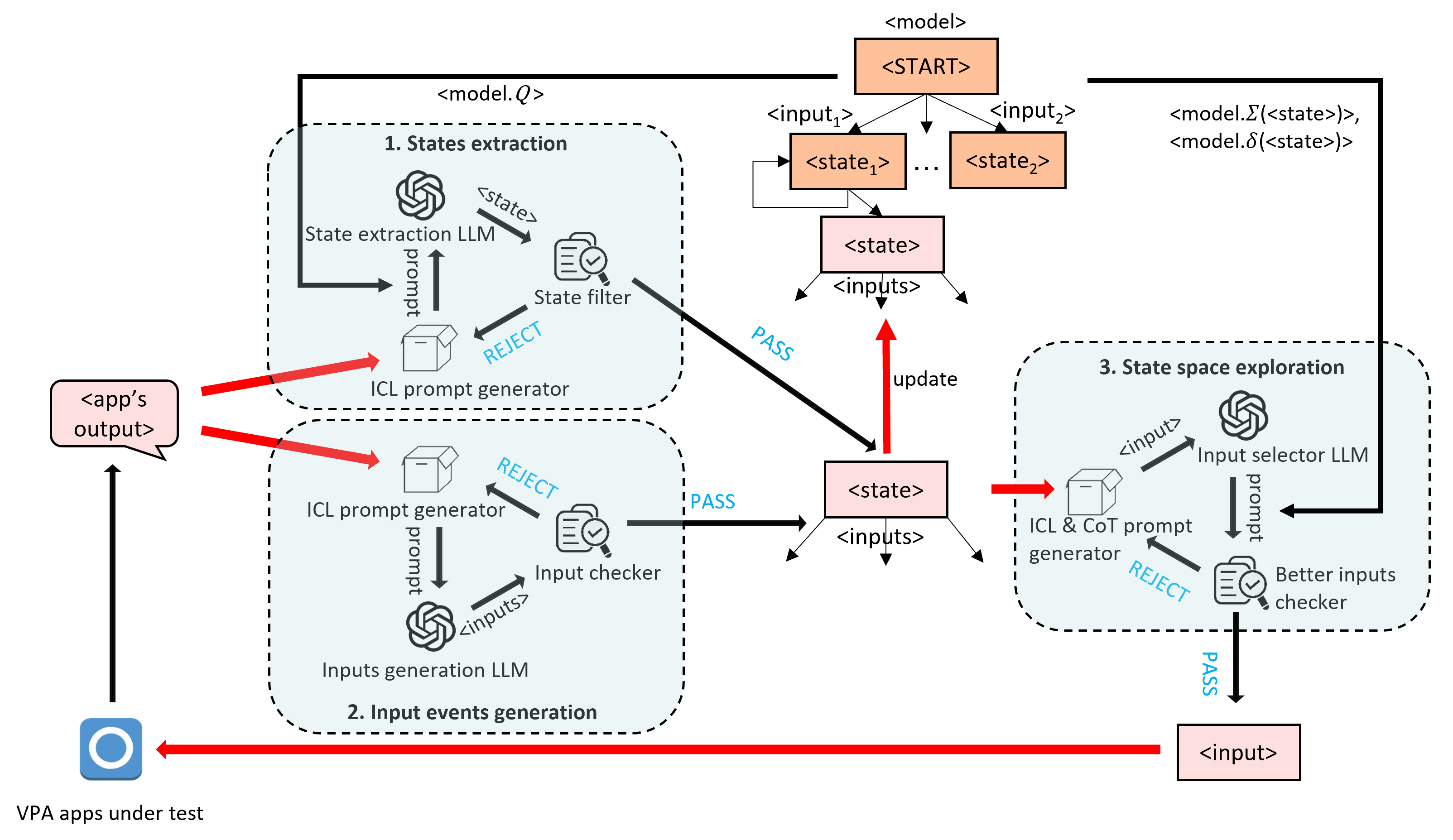}
    \caption{The framework of \sysname.}
    \label{fig:overview}
\end{figure*}

Based on these ideas, we propose {\sysname}, a model-enhanced LLM driven model-based testing method for VUI testing of VPA apps.
Figure~\ref{fig:overview} shows the framework of \sysname.
\sysname consists of three phases, and they are all performed by LLMs.
The first two phases are for model construction, including states extraction and input events generation.
In the third phase, the LLM selects an input event to explore the state space based on the information provided by the behavior model.
Since we adopt an on-the-fly model construction approach, these three phases are executed one by one repeatedly.
The main processes of these three phases are described below.

\noindent\textbf{Phase 1: States extraction.}\label{sec:state_cons}
In this phase, VPA apps' outputs and existing states in the behavior model are embedded into the prompt.
The LLM decides whether to merge the VPA apps' output with existing states or generate a new state for it.
We expect the LLM to map outputs with similar semantics to the same state.
A state filter is used to filter out mismatched states generated by the LLM.

\noindent\textbf{Phase 2: Input events generation.}
The VPA apps' real-time output is input to the LLM, which generates all possible context-related input events for this output.
We expect the input events generated by the LLM to be semantically related to the VPA apps' output and help discover meaningful new states.
An input checker is implemented to check the validation of input events according to VPA apps' feedback.

\noindent\textbf{Phase 3: State space exploration.}
The current state and current-state-related information in the behavior model are input to the LLM.
The LLM is expected to select one input event by considering factors such as the invocation frequency, historical transitions and relevance to the current state to explore the state space efficiently.
Based on the invocation frequency and history transitions, we search whether there is a better input in the input event set.
If there is one, we reject the LLM's results and ask for another input event.

Whenever we receive an output from VPA apps, we execute the first and second phases to generate states and input events.
The states and input events are used for the behavior model construction.
Subsequently, we extract information related to the current state from the behavior model and embed it to the prompt, and the LLM selects the most suitable input event at the third phase.
After that, the selected input event is fed back to VPA apps and wait for the next output.
The whole process will be continued until the time limit is reached or the VPA apps quit.
Due to the unexplainability of the LLM, we establish the feedback mechanism to check and filter out its results.
Results that do not meet our requirements are rejected, and the reasons are returned to the LLM for regenerating the results.
In the following sections, we will introduce the prompts and feedback mechanisms of these three phases respectively.

To help express the implementation of these three phases clearly, we introduce the following terms:

\begin{itemize}[leftmargin=*]
    \item \textless app's output\textgreater: the real-time VPA apps' output. It will be used to extract states. Context-related inputs are generated based on its content.
    \item \textless state\textgreater: the state extracted from \textless app's output\textgreater.
    \item \textless state$_{pre}$\textgreater: the previous explored state.
    \item \textless state$_{next}$\textgreater: the next explored state.
    \item \textless inputs\textgreater: the set of context-related inputs generated for \textless app's output\textgreater.
    \item \textless input\textgreater: the input selected by the LLM at \textless state\textgreater\ to communicate with the VPA apps.
    \item \textless input$_{pre}$\textgreater: the previous selected input.
    \item \textless model\textgreater: the behavior model.
    \item \textless model.$Q$\textgreater: the set of states in the behavior model.
    \item \textless model.$\Sigma(s)$\textgreater: the input events information of state $s$, including their invocation times.
    \item \textless model.$\delta(s)$\textgreater: the set of transition functions that start from state $s$.
\end{itemize}

\subsection{States Extraction}\label{sec:state_extrac}
Similar semantics (e.g., functionalities, purposes and context) of VPA apps can be expressed in different ways.
The LLM should merge outputs with similar semantics to one state.
For each \textless app's output\textgreater, the LLM is supposed to find a semantic similar state from \textless model.$Q$\textgreater\ or generate a new state.
For this reason, only the \textless model.$Q$\textgreater is required in this phase.
So the input of this phase includes the \textless app's output\textgreater\ and \textless model.$Q$\textgreater.

To avoid redundant results, the LLM is required to only output the \textless state\textgreater\ of the given \textless apps' output\textgreater.
To assist the LLM in better understanding this task and formalizing its outputs, we employ the in-context learning strategy.
Few shots are in the form of ``Input: \textless app's output\textgreater, \textless model.$Q$\textgreater'' and ``Output: \textless state\textgreater'' pairs.
As the LLM's results are not trustworthy, we establish a state filter to filter out mismatched states.
If a state is mismatched, we provide feedback prompts to request another state from the LLM.
The prompts of phase 1 are displayed in Table~\ref{tab:phase_1}.

\begin{table*}[!htbp]
    \centering
    \caption{The prompts of the States extraction phase.}
    \begin{tabular}{p{4.3cm}|p{12.5cm}}
    \hline
        label & prompt \\
    \hline
        *NO STATE ERROR* & The \textless state\textgreater\ is not in the state set \textless model.$Q$\textgreater. Find a semantically similar state from the state set \textless model.$Q$\textgreater\ for the sentence \textless app's output\textgreater.\\
    \hline
        *NOT MERGE SUGGESTION* & The \textless app's output\textgreater\ and \textless state\textgreater\ are not semantically similar because they have different input events.\\
    \hline
        *SHOULD MERGE SUGGESTION* & The \textless app's output\textgreater\ and \textless state\textgreater\ are semantically similar. \\
    \hline
        *LONG PROMPT* & *MAP INSTRUCTION* + *FEW SHOTS* + \textless app's output\textgreater\ + \textless model.$Q$\textgreater\\
    \hline
        *SHORT PROMPT* & \textless app's output\textgreater\ + \textless model.$Q$\textgreater \\
    \hline
        *FEEDBACK PROMPT* & *NO STATE ERROR* / *NOT MERGE SUGGESTION* / *SHOULD MERGE SUGGESTION*\\
    \hline
    \end{tabular}
    \label{tab:phase_1}
\end{table*}

When we first use the LLM for states extraction, we use *LONG PROMPT*.
In *LONG PROMPT*, we instruct the LLM to map semantically similar outputs to one states in the behavior model (labeled as *MAP INSTRUCTION*). 
Few shots are provided for LLMs to understand the state extraction task (labeled as *FEW SHOTS*).
Subsequently, we request it to return the corresponding \textless state\textgreater\ in the \textless model.$Q$\textgreater\ for the \textless app's output\textgreater.
In other cases, we will use *SHORT PROMPT*.
*SHORT PROMPT* only includes the \textless app's output\textgreater\ and \textless model.$Q$\textgreater.
After *LONG PROMPT* or *SHORT PROMPT*, the LLM will generate the \textless state\textgreater\ for \textless app's output\textgreater.
If \textless state\textgreater\ is rejected by the state filter, we will return *FEEDBACK PROMPT*.

\begin{figure}[htbp]
    \centering
    \includegraphics[width=0.45\textwidth]{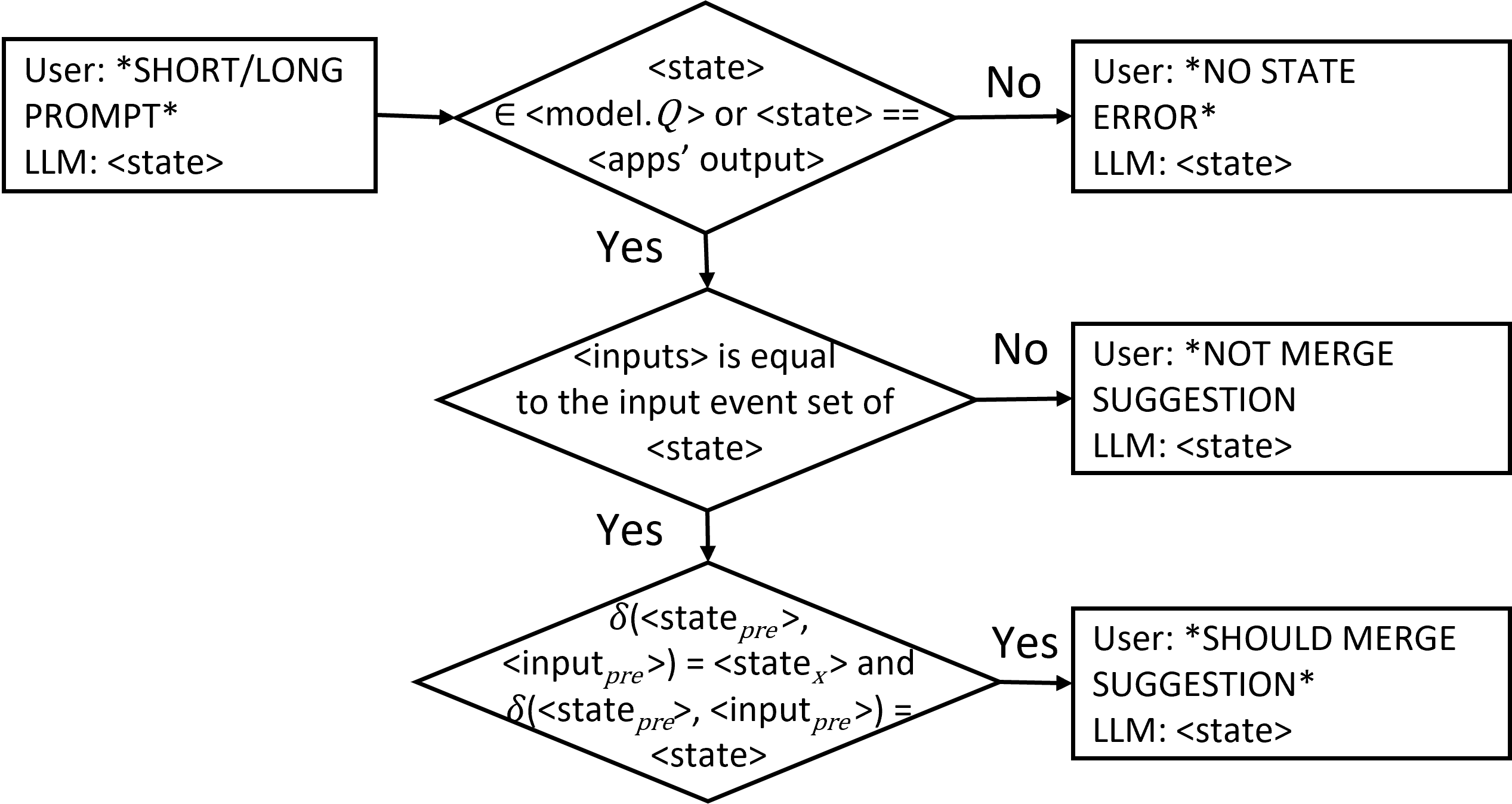}
    \caption{The workflow of the state filter.}
    \label{fig:feedback1}
\end{figure}

Figure~\ref{fig:feedback1} illustrates the state filter in the states extraction phase.
Firstly, we check whether \textless state\textgreater\ $\in$ \textless model.$Q$\textgreater\ or \textless state\textgreater\ $==$ \textless app's output\textgreater.
If neither of them is true, we return *NO STATE ERROR*.
Otherwise, we proceed to the second step of the check.
If \textless state\textgreater\ $\in$ \textless model.$Q$\textgreater, we check whether \textless state\textgreater\ and \textless app's output\textgreater\ have the same input events (see section~\ref{sec:input_gen} for the generation of \textless inputs\textgreater).
If they have different input events, we return *NOT MERGE SUGGESTION*, otherwise we move to the third step.
If \textless state\textgreater\ $==$ \textless app's output\textgreater, we find whether there exists a \textless state$_x$\textgreater\ in \textless model.$Q$\textgreater\ that satisfies the transition function $\delta$(\textless state$_{pre}$\textgreater, \textless input$_{pre}$\textgreater) = \textless state$_x$\textgreater\ and $\delta$(\textless state$_{pre}$\textgreater, \textless input$_{pre}$\textgreater) = \textless state\textgreater.
If such a \textless state$_x$\textgreater\ can be found, we consider that \textless state\textgreater\ should be merged to \textless state$_x$\textgreater.
So we return *SHOULD MERGE SUGGESTION*.


\subsection{Input Events Generation}\label{sec:input_gen}
In section~\ref{sec:state_cons}, the \textless state\textgreater\ for the \textless app's output\textgreater\ is extracted.
To further explore VPA apps' behavior, context related inputs should be generated.
Each state has its independent context related input event set, as we consider different states as different contexts.
To ensure the context relevance, the LLM is also used in this phase.
The \textless inputs\textgreater\ generated for the \textless app's output\textgreater\ is also the input event set of \textless state\textgreater.

VPA apps expect users to give short and simple inputs, but LLMs tend to generate long and redundant inputs, which most VPA apps cannot understand.
To solve this problem, we offer few shots that include five types of VPA apps' outputs~(i.e., yes-no question, selection question, instruction question, Wh question and mixed question~\cite{skillExplorer}).
For the mixed question, we summarize three most common patterns, they are instruction + selection question, Wh + selection question and yes-no + selection question.
We provide at least one example for each type of questions in the few shots.
They are in the form of ``Input: \textless apps' output\textgreater'' and ``Output: \textless inputs\textgreater'' pairs.
In addition, we set an input checker to check the validation of the input events.
The \textless state$_{next}$\textgreater\ is used to judge whether the input events generated by the LLM are context related.
If \textless state$_{next}$\textgreater\ is equal to \textless state\textgreater\ or expresses confusion, we feedback the information to request other \textless inputs\textgreater.
The prompts are displayed in Table~\ref{tab:phase_2}.

\begin{table*}[!htbp]
    \centering
    \caption{The prompts of the Input events generation phase.}
    \begin{tabular}{p{4.3cm}|p{13cm}}
    \hline
        label & prompt \\
    \hline
        *EMPTY ERROR* & The output should be a non-empty python list of the possible non-empty responses to the sentence \textless app's  output\textgreater. \\
    \hline
        *INVALID SUGGESTION* & \textless input\textgreater\ is not a valid response for the sentence \textless app's output\textgreater. The output should be a python list of *RULES*.\\
    \hline
        *RULES* & \makecell[l]{phases after ``say'' or ``ask'' (instruction question~\cite{skillExplorer})\\
        the conjunctions linked by ``and'', ``or'' and ``,''. (selection question~\cite{skillExplorer})\\
	``yes'' and ``no'' (yes-no question~\cite{skillExplorer})\\
	nouns related to \textless none\textgreater (What \textless noun\textgreater\ question)\\
	related to \textless state\textgreater (other questions)}\\
    \hline
        *LONG PROMPT* & *FEW SHOTS* + \textless app's output\textgreater \\
    \hline
        *SHORT PROMPT* & \textless app's output\textgreater\\
    \hline
        *FEEDBACK PROMPT* & *EMPTY ERROR* / *INVALID SUGGESTION* \\
    \hline
    \end{tabular}
    \label{tab:phase_2}
\end{table*}

\begin{figure}[htbp]
    \centering
    \includegraphics[width=0.45\textwidth]{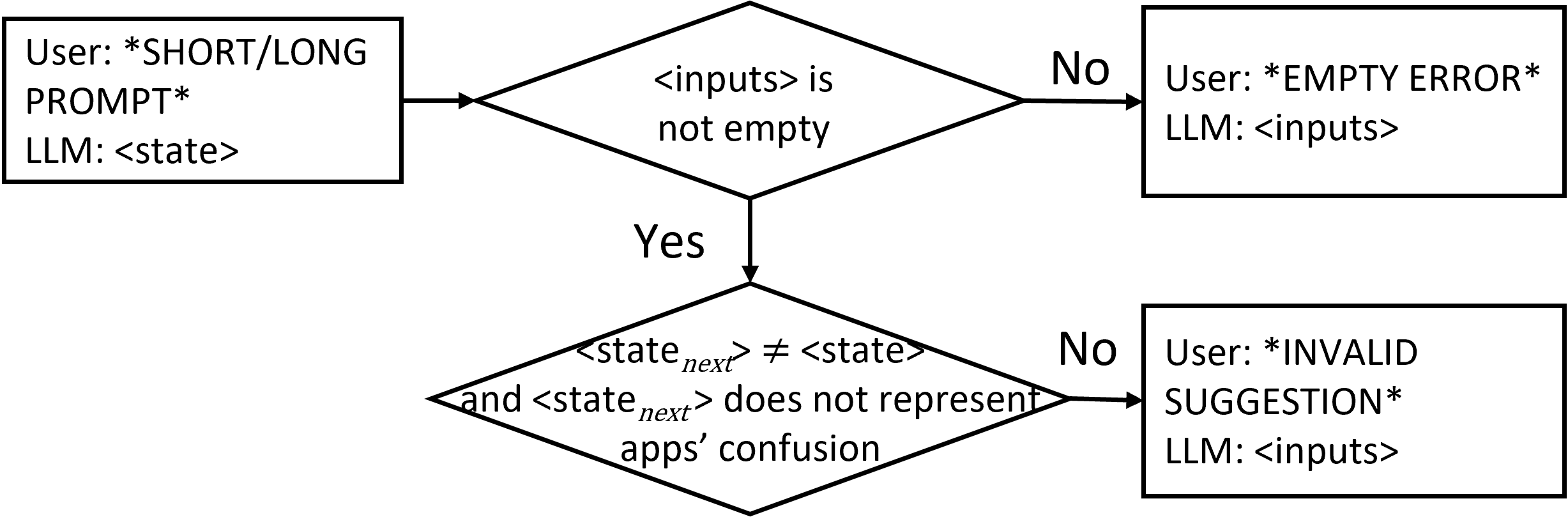}
    \caption{The workflow of the input checker.}
    \label{fig:feedback2}
\end{figure}

When we ask the LLM to generate input events for the first time, we use *LONG PROMPT*, which provides *FEW SHOTS* and instructs the LLM to find \textless inputs\textgreater\ to the \textless app's output\textgreater.
In other cases, we use *SHORT PROMPT*, which only contains the \textless app's output\textgreater.
After \textless input\textgreater\ from \textless inputs\textgreater\ is selected (see Section~\ref{sec:exploration}) and sent to the VPA app, the app will soonly give another output.
Based on the content of that output, we judge the validity of \textless input\textgreater.
Figure~\ref{fig:feedback2} illustrates the workflow of the input checker.

Firstly, we check whether \textless inputs\textgreater\ is empty.
If it is, we will return *EMPTY ERROR*.
If any input event \textless input\textgreater\ from the \textless inputs\textgreater\ is given to the VPA app and the next state \textless state$_{next}$\textgreater\ $==$ \textless state\textgreater\ or \textless state$_{next}$\textgreater\ expresses apps' confusion, \textless input\textgreater\ is considered as an invalid input event.
In this case, we will return *INVALID SUGGESTION*.

\subsection{State Space Exploration}\label{sec:exploration}
The aim of this phase is to efficiently explore the state space based on the information provided by the behavior model.
This is done by finding an input event that is most likely to discover new states (i.e., functionalities) at each state. 
It is a decision-making problem considering factors such as invocation frequency, historical transitions, and relevance to the current state based on the behavior model~(essentially a graph).
Due to the fact that LLMs have developed their abilities in understanding graphs~\cite{GPT4}, and prompt engineering techniques like chain-of-thought can improve the LLM's explainability and capability to handle reasoning tasks~\cite{cot}, the LLM is used for the state space exploration.

In the previous two phases, we extract the \textless state\textgreater\ and generate the \textless inputs\textgreater\ for the \textless apps' outputs\textgreater.
They are used to update the behavior model.
The model information is then used to guide the state space exploration.
For this reason, the input of this step includes the \textless state\textgreater\ and the \textless state\textgreater\ related information in the \textless model\textgreater.
The \textless state\textgreater\ related information includes the \textless model.$\delta($\textless state\textgreater$)$\textgreater\ and the \textless model.$\Sigma($\textless state\textgreater$)$\textgreater\ (invocation times of each input is updated after it is sent to the app).

To improve the LLM's capability of this decision-making task, we employ a strategy combining in-context learning and chain-of-thought.
We prompt the LLM to think step-by-step and show its thinking process.
In step 1, the LLM is asked to remove the input events that lead to duplicate or wrong state from the historical transitions.
In step 2, the LLM finds a never-invoked input event that is most context related.
In step 3, the LLM finally chooses one input event from the never-invoked context-related input event in step2 and the invoked and valid (i.e., does not lead to a state that is same as before or represent apps' confusion) input event.
Few shots are provided in the form of ``Input: \textless state\textgreater, \textless model.$\delta($\textless state\textgreater$)$\textgreater, \textless model.$\Sigma($\textless state\textgreater$)$\textgreater'', ``Thought: step1: xxx, step2: xxx, step3: xxx'' and ``Output: \textless input\textgreater'' triplets.
The LLM is expected to output its thinking process along with the selected \textless input\textgreater.
Similarly, the \textless input\textgreater\ given by the LLM will be evaluated and the feedback will be returned.
The prompts in this phase are displayed in Table~\ref{tab:phase_3}.

\begin{table*}[htbp]
    \centering
    \caption{The prompts of the State space exploration phase.}
    \begin{tabular}{p{4.3cm}|p{13cm}}
    \hline
        label & prompt \\
    \hline
        *NO INPUT ERROR* & \textless input\textgreater\ is not in the given input event set \textless inputs\textgreater. Please choose another input event from the input event set \textless inputs\textgreater.\\
    \hline
        *BETTER INPUT SUGGESTION* & Choosing the input \textless input$_x$\textgreater\ might be better than the input \textless input\textgreater. Please choose another input event from the input event set \textless inputs\textgreater.\\
    \hline
        *LONG PROMPT* & *MODEL DESCRIPTION* + *STEP-BY-STEP* + *FEW SHOTS* + \textless state\textgreater\ + \textless model.$\delta($\textless state\textgreater$)$\textgreater\ + \textless model.$\Sigma($\textless state\textgreater$)$\textgreater \\
    \hline
        *SHORT PROMPT* & \textless state\textgreater\ + \textless model.$\delta($\textless state\textgreater$)$\textgreater\ + \textless model.$\Sigma($\textless state\textgreater$)$\textgreater \\
    \hline
        *FEEDBACK PROMPT* & *NO INPUT ERROR* / *BETTER INPUT SUGGESTION* \\
    \hline
    \end{tabular}
    \label{tab:phase_3}
\end{table*}

\begin{figure}[htbp]
    \centering
    \includegraphics[width=0.45\textwidth]{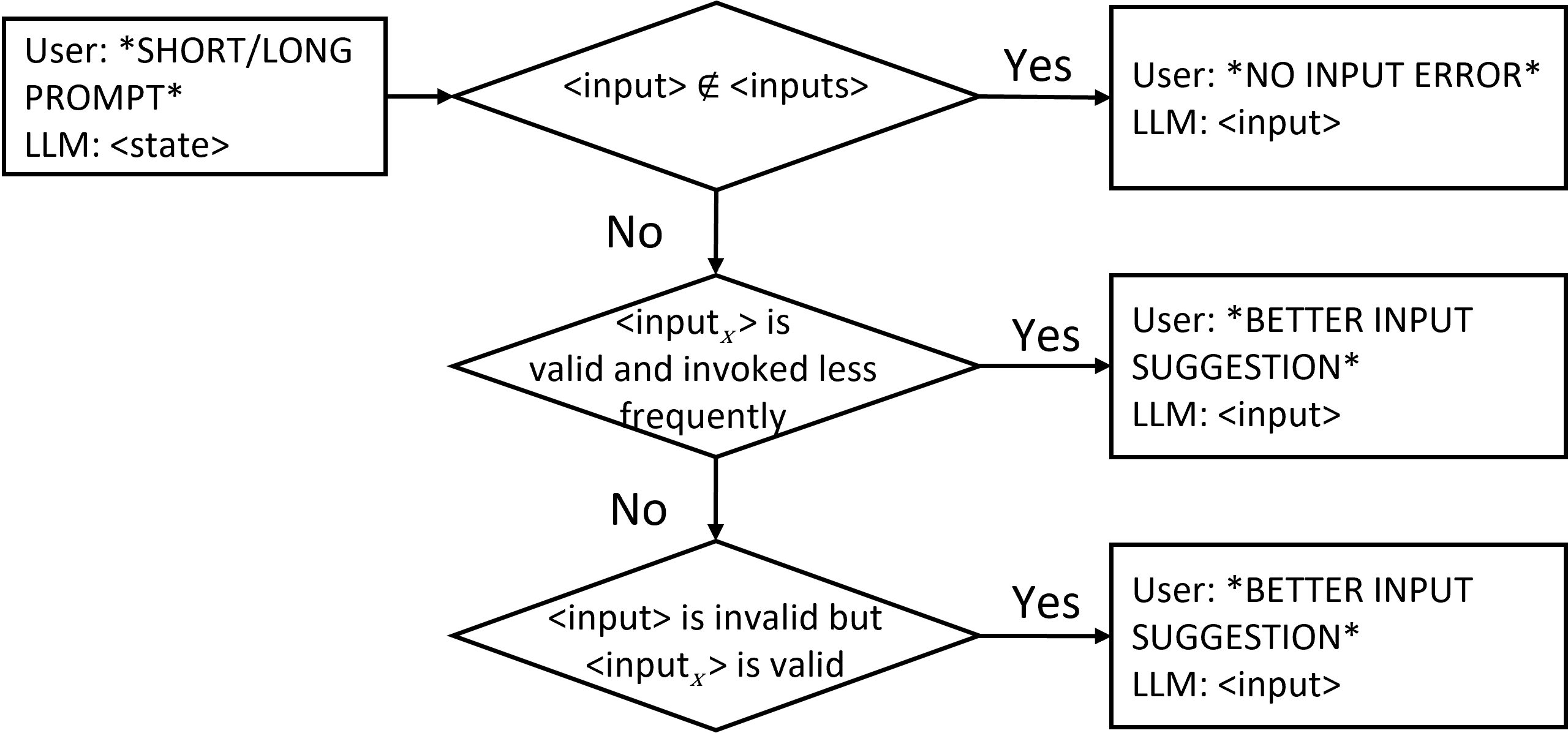}
    \caption{The workflow of better input checker.}
    \label{fig:feedback3}
\end{figure}

The *LONG PROMPT* is used for the first time.
*LONG PROMPT* initially outlines the composition and representation of the behavior model (labeled as *MODEL DESCRIPTION*).
Then, it offers step-by-step guide of the reasoning process (labeled as *STEP-BY-STEP*).
Meanwhile, few shots with the thinking process (labeled as *FEW SHOTS*) are provided.
Finally, the LLM is asked to select an \textless input\textgreater\ from the \textless inputs\textgreater\ to discover new states based on historical transitions in \textless model.$\delta($\textless state\textgreater$)$\textgreater, invocation frequency in \textless model.$\Sigma($\textless state\textgreater$)$\textgreater\ and relevance to \textless state\textgreater.
In other cases, we will use *SHORT PROMPT*, which only contains \textless state\textgreater, \textless model.$\delta($\textless state\textgreater$)$\textgreater\ and \textless model.$\Sigma($\textless state\textgreater$)$\textgreater.
After the LLM selects the \textless input\textgreater, we evaluate it by finding whether there is a probably better input event and return the *FEEDBACK PROMPT*.
Figure~\ref{fig:feedback3} illustrates the process of better inputs checker that evaluates the \textless input\textgreater\ and return different *FEEDBACK PROMPT* in the third phase.

Firstly, the better input checker checks if \textless input\textgreater\ $\in$ \textless inputs\textgreater.
If not, we return *NO INPUT ERROR*.
Otherwise, it determines whether there is a better input event \textless input$_x$\textgreater\ compared with \textless input\textgreater\ based on the invocation frequency and history transitions.
If \textless input$_x$\textgreater\ is valid and invoked less frequently than \textless input\textgreater, then \textless input$_x$\textgreater\ is better than \textless input\textgreater.
If \textless input\textgreater\ is invalid but \textless input$_x$\textgreater\ is valid, then \textless input$_x$\textgreater\ is also a better choice.
In both cases, we return *BETTER INPUT SUGGESTION*.
The \textless input\textgreater\ that passes the above checks is sent to the VPA app.

%% file: ICSE2025/evaluation.tex
\section{Evaluation}
We implement \sysname based on GPT-4~\cite{GPT4} and analyze its coverage and efficiency.
The performance of \sysname is compared with the state-of-the-art model-based VUI testing method Vitas~\cite{vitas}.
Besides, chatbot-style testers are classic VPA apps testing approach, but Vitas was evaluated to outperform traditional chatbot-style testers in coverage and efficiency.
However, with the development of LLMs, LLMs as chatbots may have stronger VPA apps testing abilities, so GPT4(chatbot) is also set as a baseline.
Additionally, we conduct ablation experiments to assess the contribution of \sysname's each phase to the final state space coverage.
We also implement \sysname on Llama2-70b-chat~\cite{Llama2-70b} and evaluate \sysname's applicability on different LLMs.
Finally, we conduct a large-scale testing on Alexa skills to evaluate \sysname's generality~\cite{skills}.

\subsection{Settings}
\noindent \textbf{Dataset}:
We use the large scale dataset of Vitas~\cite{dataset_vitas} as our basic dataset.
From this dataset, we filter out skills with no ratings.
Then, we roughly confirm 4,000 skills with consistent behavior to form the large-scale dataset.
These 4,000 skills cover all categories on the Amazon skills website. 
For the use of conducting an intensive evaluation, we also build a benchmark with 50 Alexa skills.
These 50 skills are checked to be stable and available.

\noindent\textbf{Baselines}:
We compare \sysname with two baselines, as shown in table~\ref{tab:baseline}.
The simulator provided by Amazon\cite{simulator} is used as our testing platform.
The evaluation was conducted on the Ubuntu 18.04.4 machines with AMD EPYC 7702P 64-Core Processor CPU@1.996GHz and 4GB RAM. 

\noindent\textbf{Coverage metrics}:
VPA apps are not open source, so the ground truth of the entire state space of certain VPA apps cannot be acquired in advance.
Furthermore, as \sysname merges states with similar semantics to avoid repeated testing while Vitas does not, we call the states generated by \sysname as semantic states, while the ones discovered by Vitas as sentence states in the evaluation.
Consequently, to ensure a uniform measurement, we use \sysname to process the states discovered by Vitas, and merge them to semantic states correspondingly. 
Then, we use the number of the unique semantic states achieved by \sysname and all the baselines used in certain evaluations as the total state space for each evaluation respectively for a fair comparison.

\begin{table}[!htbp]
    \centering
    \caption{Two baselines to compare with \sysname.}
    \begin{tabular}{p{1cm}|p{6.5cm}}
        \hline
        baseline & description \\
        \hline
        Vitas & Vitas is the state-of-the-art model-based testing framework for VPA apps. Vitas extracts states and generates input events through simple NLP rules and explores the state space by managing weights.\\
        \hline
        GPT4 (chatbot) & The GPT4 (chatbot) method directly uses GPT-4 as a chatbot by feeding the VPA apps' outputs to GPT-4 and returning GPT-4's results to VPA apps. No special prompts or guidance are used in this method.\\
        \hline
    \end{tabular}
    \label{tab:baseline}
\end{table}

\subsection{Evaluation of \sysname}
We aim to address the following research questions:

\noindent\textbf{RQ1}: How does the semantic state coverage and efficiency improve when using GPT-4 to enhance the model construction and exploration?



\noindent\textbf{RQ2}: Do all phases in \sysname contribute to the state exploration of VPA apps?

\noindent\textbf{RQ3}: How effective is \sysname's framework when applied to other LLMs?

\noindent\textbf{RQ4}: How is the coverage rate of \sysname on all types of skills compared with Vitas?


\subsubsection{Study1: Coverage and efficiency}
We set the time limit as 10 minutes for \sysname to test each skill.
The baselines are allowed to test skills using the same interaction rounds~(an input and an output form an interaction round) as \sysname.
Firstly, we compare the sentence states and semantic states achieved by \sysname and the baselines.
Then, we compare their average semantic state coverage with interaction rounds.

\begin{figure}[!htbp]
    \centering
    \includegraphics[width=0.4\textwidth]{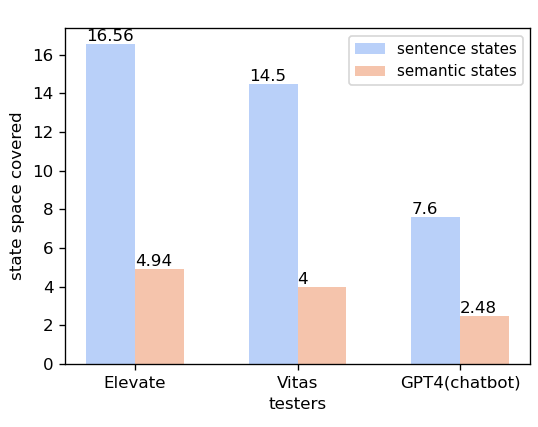}
    \caption{The comparison of the sentence states with semantic states achieved by \sysname and baselines.}
    \label{fig:comp}
\end{figure}

Figure~\ref{fig:comp} shows the sentence states and semantic states maintained by \sysname and baselines.
It suggests that the sentence states can be greatly compressed when semantic information is considered.
\sysname merges outputs with similar semantics to one state for testing, which greatly reduces the original state space.
In addition, \sysname achieves more sentence and semantic states than the baselines.

\begin{figure}[!htbp]
    \centering
    \includegraphics[width=0.45\textwidth]{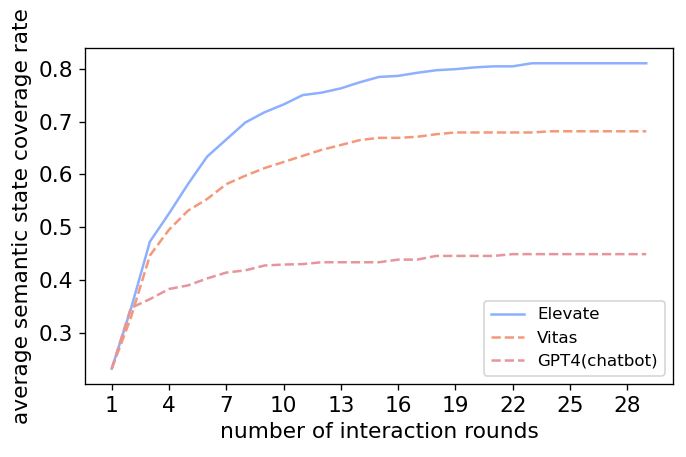}
    \caption{The average semantic state coverage rate with interaction rounds of \sysname and baselines.}
    \label{fig:study1}
\end{figure}

In order to evaluate \sysname's coverage ability along with the efficiency, we calculate the average semantic state coverage of \sysname and baselines on the benchmark of varying interaction rounds in figure~\ref{fig:study1}.
The horizontal axis represents the average semantic state space rate, while the vertical axis denotes the number of interaction rounds.
When the interactions go deeper, the advantage of \sysname over Vitas and GPT4(chatbot) is more evident.
After only 3 rounds of interactions, \sysname shows its leading exploration efficiency and stays ahead until the end.
Finally, \sysname can achieve over 80\% of average semantic state coverage after only 20 rounds of interactions, while Vitas and GPT4(chatbot) can only achieves a final coverage of 68\% and 45\% respectively.

Among the baselines, the traditional model-based tester Vitas has relatively higher performance.
However, Vitas did not exploit the semantic information during VUI testing to help the model construction and exploration, so it lags behind \sysname in terms of semantic state coverage.
Although GPT-4 is a strong LLM, directly using it as a chatbot for VPA apps testing performs worse than Vitas.
GPT4(chatbot) lacks the guidance for state space coverage, which prevents it from discovering deep states.
Enhanced with \sysname, the LLM's performance in semantic state coverage is greatly improved.

\begin{framed}
    \noindent\textbf{Answers to RQ1}:
    The sentence states can be greatly reduced when semantic information is considered.
    Compared with baselines, \sysname achieves more sentence and semantic states.
    With the increase of interaction rounds, \sysname shows evident advantage of semantic state coverage and efficiency compared with Vitas and GPT4(chatbot).
\end{framed}

\subsubsection{Study2: Ablation Studies}
To validate the rationality of prompting the LLM and returning the feedback at each phase, we conduct an ablation study.
In ``w/o States extraction'' (Section~\ref{sec:state_extrac}), ``w/o Input events generation'' (Section~\ref{sec:input_gen}) and ``w/o State space exploration'' (Section~\ref{sec:exploration}), we remove the entire *FEEDBACK PROMPT*, and the in-context learning, chain-of-thought and behavior model information of the corresponding phase in the *LONG PROMPT*.
We then let them test the benchmark using the same interaction rounds as \sysname and compare their performance on the average semantic state coverage rate.

\begin{figure}[!htbp]
    \centering
    \includegraphics[width=0.45\textwidth]{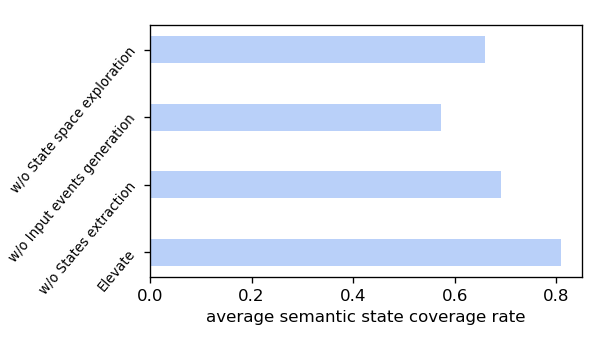}
    \caption{The comparison of semantic state coverage rate between \sysname, w/o States extraction, w/o Input events generation and w/o State space exploration.}
    \label{fig:ablation_1}
\end{figure}

Figure~\ref{fig:ablation_1} shows the average semantic state coverage rate of \sysname, w/o States extraction, w/o Input events generation and w/o State space exploration on the benchmark.
The results prove that the elimination of any phase could lead to a decrease in state space coverage.
Among them, removing the Input events generation phase has the largest impact on the final coverage, as the original input events generated by the LLM are commonly misunderstood by VPA apps.
Eliminating the w/o State space exploration phase also influences the performance.
That is because the behavior model information and chain-of-thought strategy provides the guidance for LLMs to explore efficiently.
Without the States extraction phase, the semantic state space is largely redundant, resulting in repeated tests of semantically similar states.


\begin{framed}
    \noindent\textbf{Answers to RQ2}:
    After carrying out the ablation study on \sysname's three phases, we find that each of \sysname's three phases contribute to the overall semantic state coverage rate.
    Removing the input events generation phase has the greatest impact on the final coverage rate.
\end{framed}

\subsubsection{Study3: Applicability}
We implement \sysname on Llama2-70b-chat~\cite{Llama2-70b}, referred to as \sysname-Llama2-70b-chat, to evaluate the performance of \sysname when it is implemented by other LLMs.
As a comparison, we also use Llama2-70b-chat as a chatbot to test VPA apps, and label it as Llama2-70b-chat(chatbot).
By comparing the average semantic state coverage rate of \sysname-Llama2-70b-chat, Vitas and Llama2-70b-chat(chatbot), we evaluate the applicability of \sysname.
Similarly, \sysname-Llama2-70b-chat tests skills in the benchmark for 10 minutes.
Then, Vitas and Llama2-70b-chat(chatbot) tests the benchmark using the same interaction rounds as \sysname-Llama2-70b-chat.

\begin{figure}[!htbp]
    \centering
    \includegraphics[width=0.44\textwidth]{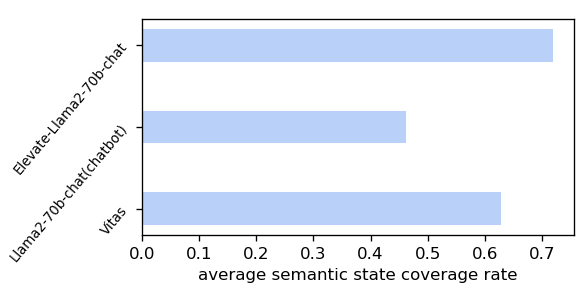}
    \caption{The comparison of semantic state coverage rate between \sysname-Llama2-70b-chat, Vitas and Llama2-70b-chat(chatbot).}
    \label{fig:applicability}
\end{figure}

Figure~\ref{fig:applicability} shows that \sysname-Llama2-70b-chat outperforms Vitas and Llama2-70b-chat(chatbot) on the average semantic state coverage rate.
\sysname's ability can be influenced by the LLM on which it is implemented on, but the result shows that \sysname-Llama2-70b-chat still has an advantage over the SOTA tester Vitas.
Besides, \sysname increases Llama2-70b-chat's coverage of VPA apps' state space by about 30\%.
Overall, \sysname's framework is applicable to other LLMs.

\begin{framed}
    \noindent\textbf{Answers to RQ3}:
    We implement the \sysname framework on Llama2-70b-chat (e.g., \sysname-Llama2-70b-chat) and compare it with Vitas and Llama2-70b-chat(chatbot).
    \sysname-Llama2-70b-chat has an advantage over Vitas and Llama2-70b-chat(chatbot) in the average semantic state coverage rate.
    Additionally, \sysname increases Llama2-70b-chat's coverage of VPA apps' state space by about 30\%.
    Therefore, the \sysname framework is applicable to other LLMs.
\end{framed}

\subsubsection{Study4: Generality}
In the preceding studies, we evaluate the coverage and efficiency capabilities of \sysname on the small scale benchmark.
In this study, we use \sysname to test 4,000 skills in the large-scale dataset.
By comparing its average coverage rate with Vitas in all categories, we evaluate its ability to test skills with various functionalities.
As the cove The total coverage is set as the union of the unique coverage achieved by Vitas and \sysname.



\begin{figure*}[!htbp]
    \centering
    \includegraphics[width = 0.98\textwidth]{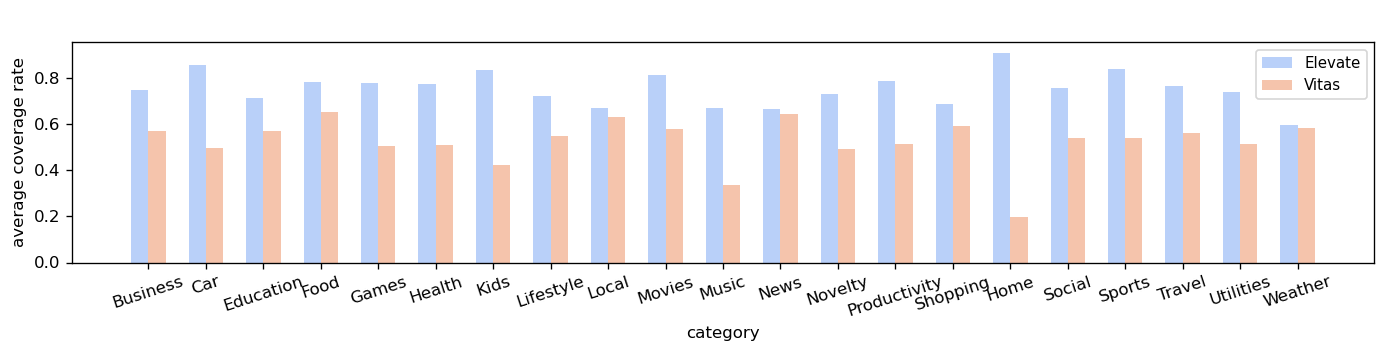}
    \caption{Average semantic state coverage rate with different categories on the large scale dataset compared with Vitas}
    \label{fig:large-scale}
\end{figure*}

The average semantic state coverage rate with different categories compared with Vitas on the large scale dataset is shown in figure~\ref{fig:large-scale}.
The results demonstrate that \sysname can achieve over 15\% of higher semantic state coverage rate in most categories compared with Vitas.
It proves \sysname's ability to test skills with different behavior.
\sysname is enhanced with LLMs, which are trained on massive amounts of data, enabling their abilities to handle a wide variety of VPA apps.
As a comparison, Vitas is designed with fixed patterns to process all types of VPA apps.
Consequently, Vitas may lack generality when applied to specific VPA apps.

\begin{framed}
    \noindent\textbf{Answers to RQ4}:
    Compared with Vitas, \sysname demonstrates a 15\% of higher semantic state coverage rate on most categories of skills.
    The results prove the generality of \sysname on testing various VPA apps.
\end{framed}

%% file: ICSE2025/discussion.tex
\section{Discussion}

\subsection{\sysname's limitations}
\sysname's limitations primarily lie in the large language model.
Firstly, although the LLMs can achieve good results, their outputs are non-deterministic.
Hence, the performance may vary with each test.
Secondly, the thinking process of the LLM is not always accurate.
As we introduce the chain-of-thought method in the third phase, the LLM will output its thinking process.
While chain-of-thought can enhance coverage and efficiency, the thinking process of the LLM is not always right and we cannot confirm whether the LLM is actually thinking as we expected.
Lastly, in rare cases, the LLM may not rectify the results even after multiple rounds of feedback prompts.
In such instances, we consider that our feedback strategy cannot steer the LLM out of its hallucination and we resort to generate states and input events based on simple rules.

%% file: ICSE2025/related-work.tex
\section{Related Work}
\noindent\textbf{VPA apps Testing}:
Several studies have been conducted to test quality, privacy or security related problems on VP apps~\cite{skillExplorer,verhealth,vitas,skilldetective,VUI_UPSET_1,VUI_UPSET_2,skillscanner}.
SkillExplorer~\cite{skillExplorer}, VerHealth~\cite{verhealth} and SkillDetective~\cite{skilldetective} are chat-bot style testers that focuses on detecting skills' privacy violation behavior.
SkillExplorer and SkillDetective~\cite{skilldetective} adopt the DFS-based exploration approach.
VUI-UPSET~\cite{VUI_UPSET_1,VUI_UPSET_2} is a chat-bot style testing approach to generate correct paraphrases while detecting bugs.
Vitas~\cite{vitas} uses the model-based testing to test VPA apps' problems related to quality, privacy and security.
Despite the improvement in coverage and efficiency, it uses simple rules to construct the model and fails to consider the semantic information.
SkillScanner~\cite{skillscanner} is the first static analysis method to identify skills' policy violations at the development phase based on a dataset collected from the GitHub.
Compared with them, \sysname adopts the model-based testing approach to improve the exploration efficiency and introduces to use the LLM to supplement missing semantic information for model construction and exploration.

\noindent\textbf{Security and Privacy of VPA apps}:
Increasing number of research focuses on security and privacy issues of VPA apps~\cite{survey_s&p_1,survey_s&p_2,survey_s&p_3}.
Kumar et al. proposes the skill squatting attack~\cite{skill_squatting}.
Several searches detected the weakness of the automatic speech recognition (ASR) system, which is vulnerable to adversarial sample attacks and out-of-band signal attacks~\cite{ASR_attack_1,ASR_attack_2,ASR_attack_3,ASR_attack_4}.
Many efforts have been spent on detecting problematic privacy policies and potential privacy violating behavior~\cite{policy_1,policy_2,skillExplorer,skilldetective,skipper}.
Different from them, \sysname sought to thoroughly explore the VPA apps' behavior so that sufficient problems can be discovered.

\noindent\textbf{Large Language Model for Software Testing}:
As a booming new technology, Large Language Models are applied to many areas, including software testing.
Codet~\cite{codet} uses the LLM to automatically generate test cases for evaluating the quality of a code solution.
CodaMosa~\cite{CodaMosa} asks Codex to generate test cases when the search based software testing method reaches the bottleneck.
TitanFuzz~\cite{titanfuzz} uses LLMs to generate and mutate input DL programs for fuzzing DL libraries.
Its follow-up work, FuzzGPT~\cite{fuzzgpt}, primes LLMs to synthesize bug-triggering programs for fuzzing and shows improved bug detecting performance.
Other research focused on testing the GUI of mobile apps by generating context-related texts or human-like actions~\cite{LLM_GUI_1,LLM_GUI_2}.

%% file: ICSE2025/conclusion.tex
\section{Conclusion}
In this work, we propose \sysname, a LLM driven model-based testing framework for VPA apps.
\sysname uses the LLM for constructing the behavior model and exploring the state space to compensate for the loss of semantic information.
It extracts states from VPA apps' outputs and generates input events to these outputs by providing few-shots to LLMs.
The LLM's exploration ability is enhanced by chain-of-thought.
Moreover, \sysname sets checkers to analyze the LLM's results and uses feedback prompts to ask LLMs for adjustments.
Our experiments show that \sysname achieves higher coverage than the state-of-the-art tool Vitas and LLMs as chatbots in an efficient manner.
\sysname tests a large-scale dataset of 4,000 Alexa skills and achieves about 15\% of higher coverage rate than Vitas in all categories.